# Coherent X-ray Imaging of Stochastic Dynamics

Arnab Sarkar[a] and Allan S. Johnson[a]*

[a.]*IMDEA Nanoscience, Calle Faraday 9, 28049, Madrid, Spain.*

*allan.johnson@imdea.org

**Condensed phase systems often exhibit a mixture of deterministic and stochastic dynamics at the nanoscale which are essential to understanding their function, but can be challenging to study directly using conventional imaging methods. Coherent X-ray imaging has emerged as a powerful tool for studying both nanoscale structures and dynamics in condensed phase systems, including stochastic dynamics, but the requirement to obtain single-shot images in order to obtain freeze-frame images of the stochastic dynamics means the X-ray fluxes used must be very high, potentially destroying the samples. This prevents coherent imaging from being applied to complex systems like tracking the motion of charge carriers or domain fluctuations in quantum materials. Here we show that, by leveraging the coherence intrinsic to these methods, we can separate out the stochastic and deterministic contributions to a coherent X-ray scattering pattern, returning real space images of the deterministic contributions and the momentum spectrum of the stochastic contributions. We further show that, for several typical and important classes of fluctuations, we can return real space images of the mean fluctuations. We demonstrate this approach by numerically simulating the imaging of stochastic polaron separation following photoexcitation and by recovering the spectral properties of fluctuating domain walls. Our versatile approach will enable the direct recovery of the spatial, spectral and temporal properties of stochastic material dynamics in a wide variety of systems currently unobtainable with existing methods.**

## Introduction

X-ray microscopy is a powerful tool for imaging nanoscale structure in a wide range of biological and materials systems. The ability to return real space images with nanometric resolution, chemical sensitivity and spectroscopic information has proved indispensable in understanding catalytic system, light-harvesting systems, structural mechanics and devices[1–4]. Measuring the dynamics in such systems, however, has proven much more challenging. X-ray optics are often incompatible with relevant sample geometries, and scanning methods require highly repeatable events to build up images of dynamical processes[5]. Recently great progress has occurred in imaging dynamics using coherent lensless X-ray imaging methods, where X-ray optics are dispensed with and instead the spatial coherence of the beam is leveraged to obtain diffraction patterns which can be numerically inverted to return real space images[6]. In the case of Coherent Diffractive Imaging (CDI) this inversion is performed using iterative phase reconstruction algorithms[7,8], while in the related technique of Fourier Transform Holography (FTH), holographic reference holes add a local oscillator term and allow direct inversion via Fourier transform[9]. As full-field imaging methods both return a complete picture of the object simultaneously, and so are well suited for imaging dynamical processes as well[5].

Both FTH and CDI have recently been leveraged to image real space dynamics in quantum materials[10,11], but acquiring videos of dynamical processes remains extremely challenging. In order to form an image of a dynamical process it is necessary to capture enough photons on the detector, while ensuring that they scattered from the sample within the time window of the event. For even quite slow nanoscale processes this becomes challenging, as the X-ray fluxes required would introduce tremendous heating and because imaging detectors are generally limited to, at best, kilohertz acquisition rates. The aforementioned studies[10,11] overcame this limitation by using stroboscopic measurements, in which the dynamical process is repeatedly initiated and a weak pulse of X-rays captures a partial frame each time (a pump-probe measurement). These partial frames are then averaged to obtain sufficient signal to construct an image. The major limitation here is, of course, that it needs to be possible to repeatedly initiate the dynamical process[12]. This rules out measurement of *stochastic* processes, where the system evolves through a different path in phase space during each measurement.

Such stochastic processes are widespread at the nanoscale, where thermal or quantum effects become highly significant[13–15]. For instance quantum materials often show stochastic motion of charge carriers, vortices, or domain walls[14,16,17]. Because of the difficulty in forming real space images of such stochastic processes, fluctuations are generally studied through alternative methods that return the statistical properties[18–21]. Only recently has real space imaging been attempted using coherent methods with the methodology of coherent correlation Imaging (CCI)[22], in which partial frames are taken faster than the timescale for the fluctuations. Similar frames are then grouped until the signal-to-noise is sufficient to reconstruct real images. CCI is a major methodological advance, but still requires enough flux in order to ensure the partial frames are sufficiently complete to allow for sorting, and furthermore requires a discrete number of different fluctuation parameters in order for the grouping procedure to be meaningful. Alternatively, the advent of high-intensity free-electron lasers has enabled single-shot imaging with femtosecond pulses[23], which should allow snapshots of fluctuations, though this may not be possible in many systems due to damage concerns[7]. In systems where single-shot imaging of the fluctuations is not feasible, the best case scenario would be to return real-space information on the deterministic part of states or reaction pathways, while simultaneously returning directly the statistical properties of the fluctuations independently.

Here, we demonstrate a new method for separating the stochastic and mean (deterministic) contributions in coherent imaging methods which we dub *Coherence Isolated Diffractive Imaging* (CIDI). By introducing fluctuation-free interferometric references to a coherent scattering pattern in an FTH arrangement, we show the mean contribution is fully encoded in the coherent interferometry data. Conversely, the fluctuations are not reflected in the interferogram, allowing the mean and stochastic contributions to be separated *post-hoc* through a fully analytic algorithm. In particular CIDI allows us to recover the full momentum distribution of fluctuations as a function of external parameters like photon energy or pump-probe delay, allowing us to return important information on their size and properties. This information is not readily accessible with any existing methodology, and CIDI could have a unique role to play in understanding systems as diverse as light-harvesting complexes, superconductors, and catalytic systems.

## Separation of Coherent and Incoherent Contributions to X-ray Holography

We consider the case where a coherent electromagnetic plane-wave scatters from a finite sample and the diffraction pattern is recorded on detector, in an FTH-like arrangement. In X-ray holography a mask constrains the object to an aperture of radius $D$. Small transmissive reference holes placed at a position $> 4D$ away from the center provide a phase reference (Figure 1a), allowing retrieval of the image directly from the corresponding diffraction pattern (Figure 1b) by a simple Fourier transform (Figure 1c)[9]. In the presence of stochastic dynamics occurring faster than the exposure time, or alternatively when averaging over multiple partial frames, the scattering pattern on the detector appears as the incoherent sum of the diffraction pattern of all the configurations.

The simple intuitive idea behind CIDI is that the "static" or, more precisely, deterministic contributions lead to the same scattering pattern and fringe positions in each frame, while the stochastic portion leads to a changing fringe position between each exposure. This means the fringe visibility will be lost for the stochastic portion but remain for the deterministic portion. Because FTH is an interferometric technique, it is sensitive to interferometric fringes only and insensitive to the low-fringe contrast stochastic background[24]. By applying FTH analysis we can then separate these two components and extract the scattering pattern of the stochastic contribution alone.

**Analytic theory**

The total electric field from the sample and holography holes can be written as a sum of three components:

$$E(r,t) = x(r) + R(r) + S(r,t). \quad (1)$$

Here, $x(r)$, $R(r)$ and $S(r,t)$ are contributions from deterministic part of the object, reference hole, and stochastic part, respectively, while $r$ denotes the spatial coordinates and $t$ time. Note that this could be either be real time, in the case of an extended exposure[22], or frame number in the case of a stroboscopic measurement[10,11]. Both the static and stochastic parts are restricted to the object aperture, i.e. $x(r > D) = S(r > D) = 0$. By definition the time average of the stochastic portion $\langle S(r,t) \rangle = 0$; any non-vanishing contribution (for instance an overall decrease in transmission across the entire sample) is included in $x(r)$.

The time averaged Fourier transform of electric field squared, which is the quantity actually measured at the detector in coherent imaging experiments, can be written as:

$$\langle |f(k,t)| \rangle = \left\langle \left| FFT(E(r,t)) \right|^2 \right\rangle$$
$$= \int dt (|x(k)|^2 + |R(k)|^2$$
$$+ |S(k,t)|^2 + R(k)x^*(k)$$
$$+ R^*(k)x(k) + S(k,t)x^*(k)$$
$$+ S^*(k,t)x(k) + S(k,t)R^*(k)$$
$$+ S^*(k,t)R(k)). \quad (2)$$

$R(k)$ is the Fourier transform of the reference hole $R(r)$, $x(k)$ is the Fourier transform of the static contribution, $S(k,t)$ is the Fourier transform of $S(r,t)$, and $\langle \rangle$ denotes a time averaged quantity. The cross-correlations with the stochastic term vanish due to time averaging since $\langle S(k,t) \rangle = 0$ and the expression simplifies to:

$$\left\langle \left| FFT(E(r,t)) \right|^2 \right\rangle = |x(k)|^2 + |R(k)|^2 + |S(k)|^2$$
$$+ R(k)x^*(k) + R^*(k)x(k). \quad (3)$$

where $|S(k)|^2$ is the time averaged momentum space spectrum of the fluctuations (henceforth the fluctuation spectrum). We rearrange this expression to isolate the fluctuation spectrum as a function of the experimentally measured scattering pattern and the deterministic components:

$$|S(k)|^2 = \left\langle \left| FFT(E(r,t)) \right|^2 \right\rangle - |x(k)|^2 - |R(k)|^2$$
$$- R(k)x^*(k) - R^*(k)x(k). \quad (4)$$

To evaluate this expression, we need to find the autocorrelation terms of static part of the sample and holography hole. This can be done by applying a simple holographic analysis which extracts $R^*(k)x(k)$ directly (Figure 1c). In the limit where $R(x) = \delta(x)$, i.e. the standard holographic case[9], this also yields $x(k)$ and the full expression can be evaluated, returning $|S(k)|^2$.

A significant improvement can be obtained by considering the case where there are two holography holes[25,26] (see figure 1a), which allows us to extract all the autocorrelation terms separately without the need for the small aperture approximation. Considering electric fields from two holographic apertures $R_1(r)$ and $R_2(r)$, we can rewrite the equation 4:

$$|S(k)|^2 = \left\langle \left| FFT(E(r,t)) \right|^2 \right\rangle - |x(k)|^2 - |R_1(k)|^2$$
$$- |R_2(k)|^2 - R_1(k)x^*(k)$$
$$- R_1^*(k)x(k) - R_2(k)x^*(k)$$
$$- R_2^*(k)x(k) - R_1^*(k)R_2(k)$$
$$- R_1(k)R_2^*(k). \quad (5)$$

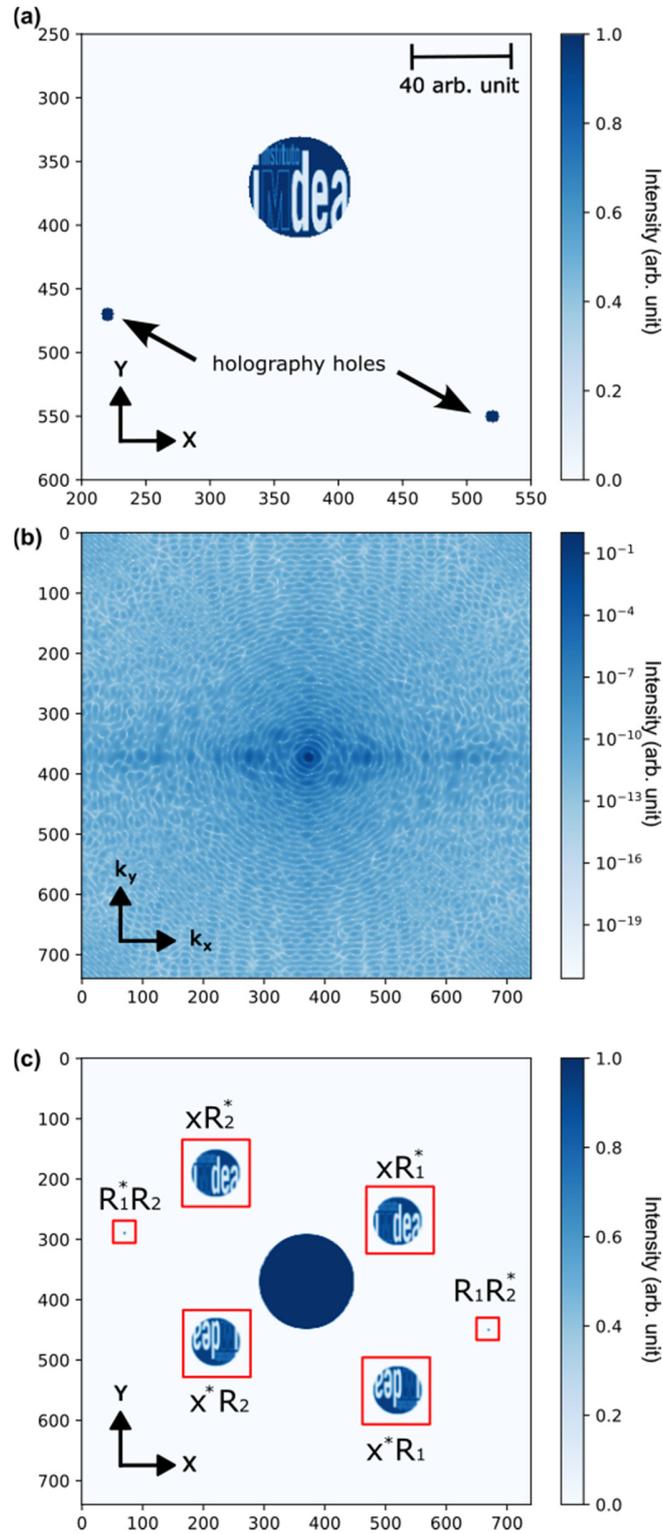

**Figure 1: Geometry for CIDI analysis.** (a) FTH-like mask geometry. A two-dimensional transmissive test object is constrained inside a circular mask with two reference holes (small blue dots) nearby providing phase references. b) The corresponding diffraction pattern that appears on the detector plane. c) Retrieved holography image obtained from direct Fourier transform of panel b. The pattern consists of all the autocorrelation terms (at the centre) and cross-correlation terms (images outside the central circle). The image is saturated to show the structure in the cross-correlation terms.

The first term of the right-hand side of this equation is the total diffraction pattern in presence of both the reference holes (Figure 1b), while the others are the various cross and autocorrelation terms. All the cross-correlation terms can be directly extracted from different parts of the Fourier transform of retrieved holography image, as indicated in Figure 1c. This then allows us to uniquely extract the autocorrelation terms as follows.

$$|x(k)|^2 = \left|\sqrt{\frac{R_1(k)x^*(k) \times R_1^*(k)x(k) \times R_2(k)x^*(k) \times R_2^*(k)x(k)}{R_1^*(k)R_2(k) \times R_1(k)R_2^*(k)}}\right|, \quad (6)$$

$$|R_i(k)|^2 = \frac{R_i(k)x^*(k) \times R_i^*(k)x(k)}{|x(k)|^2},$$

where the × operator is used to separate each pair of constituent cross-correlation terms that can be identified directly from the holography image. Thus, the fluctuation spectrum can fully be retrieved without any *apriori* assumptions.

The ability of an FTH experiment to directly return the fluctuation spectrum via CIDI analysis is the main finding of this article. While fluctuations are often studied via diffraction measurements, previous methods are unable to separate the contributions from the fluctuations and the average behaviour directly; for instance, electron diffraction studies[18,20,27] on topological defects yield the spacing between the fluctuating topological defects, rather than the scattering from the defects themselves. As we will show in the rest of this article, CIDI allows us to extract the spatial and spectral properties of fluctuations directly. Finally, we note that because the summation of the scattering patterns is incoherent (different frames) the signal levels of the stochastic and deterministic patterns both scale linearly with the number of exposures or overall signal level. All operations thereafter are linear, and so the relative signal levels of these two contributions are also linearly related to their actual contributions to the transmission of the object. Another interesting and perhaps counterintuitive feature is that while holographic imaging usually does not allow imaging of features smaller than the size of the holographic reference aperture $d$[26], the CIDI procedure isolates the fluctuation spectrum up to the maximum momentum $k$ supported by the detector or wavelength of like. The main effect of the finite size of the reference aperture is to introduce zeros in the Fourier transform $R_i(k)$ at $k = 1/d$. This introduces divergences which must be masked when extracting the autocorrelation features, but the momentum information at higher $k$ remains meaningful, unlike in FTH where it introduces a phase ambiguity. Numerically, we routinely recover meaningful information at $k$ far higher than $1/d$.

**Coherent diffractive imaging of a stochastic spectrum**

While direct access to the momentum spectrum of the fluctuations is already a powerful new capability, enabling the extraction of important information like characteristic length scales, spectral dependencies and temporal evolution, the unique geometry used in X-ray holography actually allows us to go further still. Because the fluctuations are fully contained inside the mask, they naturally fulfil the confined sample requirement for CDI reconstruction[28,29]. This gives the opportunity to, for certain types of fluctuations, retrieve the phase information of the fluctuating objects by using CDI algorithms and return real-space images of the fluctuations. This naturally depends on how well the average fluctuation spectrum $|S(k)|^2$ corresponds to the spectrum of the fluctuation, since CDI requires the spectrum in k-space corresponds to a unique object in real space – a condition which does not have to be satisfied for an incoherent averaging.

Nevertheless, there are two reasons to consider this approach. The first is that for a particular class of fluctuation – position only fluctuations – the correspondence between the average fluctuation spectrum $|S(k)|^2$ and the spectrum of each individual fluctuation *is exactly equal*. This is because fluctuations in position, for example a quasiparticle nucleating in a different region each acquisition, means only fluctuations in phase in k-space, which are lost regardless. This, in normal CDI, leads to ambiguity in absolute positioning, but here allows to exactly reconstruct such fluctuations. The second reason CDI may be considered to yield interesting insight, even in the case where exact correspondence cannot be assured, is simply because CDI is well known practically to converge to representative mean solutions in many such cases[30]. This can be seen in recent work where CDI reconstructions of fluctuating magnetic domain patterns converged to the mean pattern[22]. Similar behaviour has long been known in the very closely related problem of numerical ultrafast pulse reconstruction[24].

## Numerical Demonstrations

To verify the analytic theory, we have numerically simulated a variety of different scenarios and compare the output of the CIDI analysis to the input numerical data. In all cases we consider the FTH geometry shown in Figure 1a, and then introduce fluctuations into the central image aperture. For simplicity we consider two perfectly circular reference holes with same radius, but the analysis is robust with respect to variation in these parameters. Unless otherwise stated, we average over 5000 frames in order to realize the statistical limit for the fluctuations.

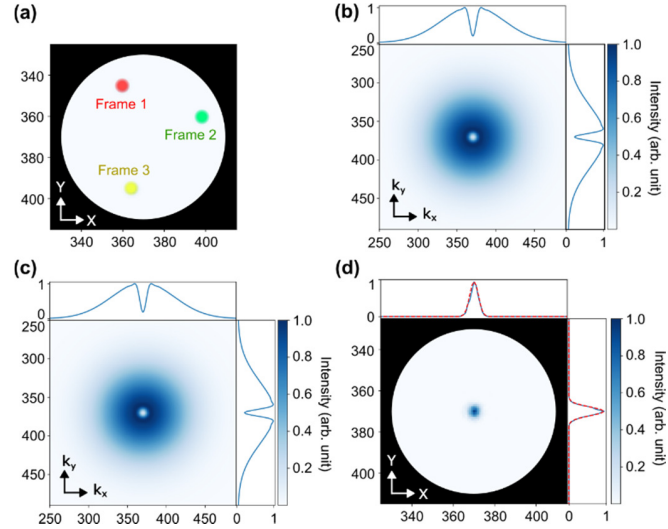

**Figure 2: CIDI of single vortices.** a) Image of a single vortex with fluctuating position in two dimension is shown for three frames. The three different locations are represented by three different colours. b) Fluctuation spectrum of a single vortex calculated directly from the fluctuations. c) Fluctuation spectrum retrieved using CIDI. d) Image of a single vortex reconstructed by applying CDI to panel c, compared with the original vortex shape (red lineouts).

**Single vortex dynamics**

As a first example, we consider a single vortex (a two-dimensional Gaussian function) appearing at an arbitrary location inside the object mask. We model the case where the stochastic dynamics of the vortex are faster than the acquisition rate, and the probability of finding a single vortex at a particular point on the sample is random and uncorrelated. In figure 2a, a representative image of three frames is shown, where different colors (red, green and yellow) represent different snapshots of the vortex with the static background removed. In this case we don't observe any spatial information for the vortex in the average image (not shown) – the effect of the vortex is just to change the average transmission of the sample. By applying the CIDI algorithm and numerically subtracting the static information, however, we isolate the momentum spectrum of the vortices as shown in Figure 2c. The fluctuation spectrum calculated directly from the introduced fluctuations, without application of the CIDI procedure, is shown in Figure 2b and matches exactly to the retrieved spectrum. The fluctuation spectrum in this case is a Gaussian distribution in momentum space with a dip at the center. This dip corresponds to the Airy disk resulting from the change in average transmission introduced by the vortices, which is reflected in $|x(k)|^2$ rather than in $|S(k)|^2$.

As the fluctuations introduced here are in position only, we can next apply CDI to the retrieved spectrum. Using the known aperture size extracted from the holography analysis[25] as our mask in the object plane, we apply 500 iterations of a modified relaxed alternating reflections algorithm[31]. As described above, we mask regions where the retrieved fluctuation spectrum diverges due to zeroes in $R_i(k)$ and allow the solution to vary freely in these regions. Good convergence is found for all reported CDI reconstructions within this article. Figure 2d shows the reconstruction fluctuation object, which matches identically with the original vortex structure (red dotted line on side panels) apart from a small DC offset corresponding to the average transmission change.

Extending the same method to extract information about multiple uncorrelated vortices with arbitrary locations is straightforward. If we have more than one vortex, the resulting interference pattern depends on the specific configuration, and the fluctuation spectrum averages over these non-identical patterns. The high-frequency fringes, corresponding to the separation between different vortices, averages away over many configurations, while the overall Gaussian structure corresponding to the structure of each individual vortex survives this process.

**Polaron and charge carrier pairs**

We next simulate a pair of particles, for instance polarons or other charge carriers[32]. We represent them as two Gaussians with equal but opposite changes in X-ray transmission, roughly corresponding to the shifts in absorption expected for a positive and negative charge carrier[32]. This is very general situation, and we simulate a variety of different configurations to understand the utility of the method under different scenarios. Initially we simulate a pair of polarons with opposite sign separated by a fixed distance along one spatial axis corresponding to, for instance, a snapshot of ballistic motion following photoexcitation. Figure 3a shows three snapshot frames of the polaron pair subtracting the static background of the object. At each frame, the polaron pair appears at an arbitrary position but keeping the same relative distance and orientation. We do not, however, constrain both polarons to appear within the object aperture, and so there are frames in which only one polaron contributes to the X-ray scattering. After applying the CIDI algorithm we recover the scattering pattern shown in figure 3b. The fringe pattern is immediately reminiscent of that from a double slit experiment, with the fringes encoding the separation of the sources and their relative orientation. Note that here, because the net contribution of the positive and negative polarons to the transmission is zero, the Airy disk feature at $k = 0$ does not appear. We can then apply CDI to the recovered fringe pattern in Figure 3b following the same procedure as for the single vortices. We successfully retrieve the amplitude and phase (figure 3c and d) of the polaron pairs, unambiguously showing the pattern results from a pair of sources with opposite sign, while also successfully recovering their size. We further consider the case of a pump-probe experiment in which at different delay times the polaron separation increases (I-V). As can be seen we clearly resolved the increasing separation, and CIDI can be used to track polaron motion in stroboscopic measurements.

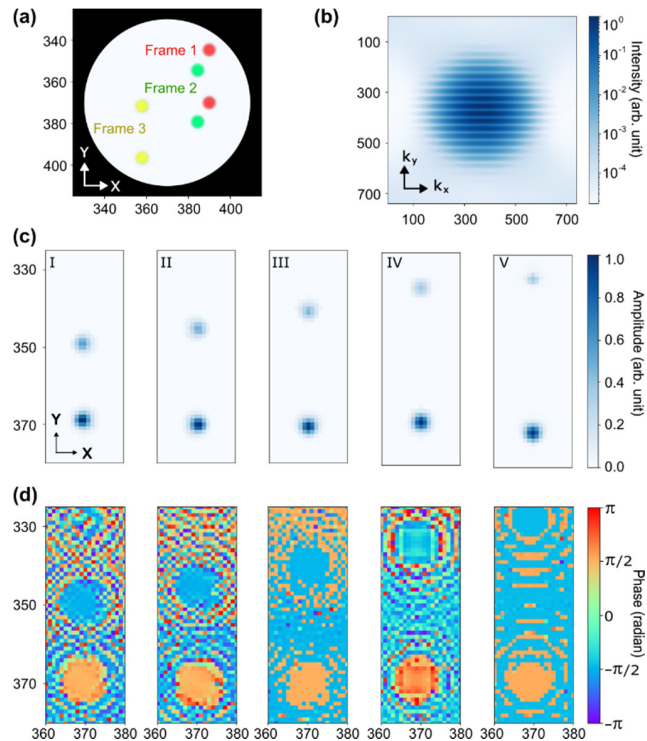

**Figure 3: CIDI of oriented polaron pairs**. a) An image of a pair of polarons separated along Y axis with fluctuating mean position in two dimension is shown for three frames. The three different locations are represented by three different colours. b) Fluctuation spectrum of the pair of polarons retrieved using CIDI. c) Amplitude and d) phase image of a pair of polaron moving away from each other (from I to V) retrieved using CDI. We are able to return the size, seperation, and relative phase of both polarons.

It is, however, fairly unlikely that most charge carriers or other paired particles will propagate along only one axis. Thus we simulated the case where the relative distance remains same but both the position and orientation of the polaron pairs is random for each frame, as shown in Figure 4a. This is naturally a much more challenging scenario – it no longer corresponds to the case of a position only fluctuation, and so the fluctuation spectrum no longer maps to the momentum spectrum of a singular object as in the previous cases. In particular, the clear fringe pattern observed in Figure 3b is blurred out by rotational averaging and we observe a series of rings in the fluctuation spectrum as shown in Figure 4b. Nevertheless, clear oscillations in the radial momentum are still observable, encoding the relative separation of the polarons. We apply CDI reconstruction methods to this retrieved pattern and find that we are able to converge to a clear image, a central peak within a large ring (Figure 4c and d). The radius of ring matches the separation of the polaron pairs, while the width of the ring and of the central aperture encodes the size of the polarons. The phase difference between the inner and outer ring furthermore matches the phase shift between the two polarons. Thus we are able to return a range of quantitative information even in this significantly more challenging scenario. We briefly note that in the case where the average separation of the polarons is also statistical, for instance in a diffusive scenario, the separation of the polarons and their relative sizes would be convolved, preventing such direct assignment, though properties such as velocity and relative phase would still be resolvable.

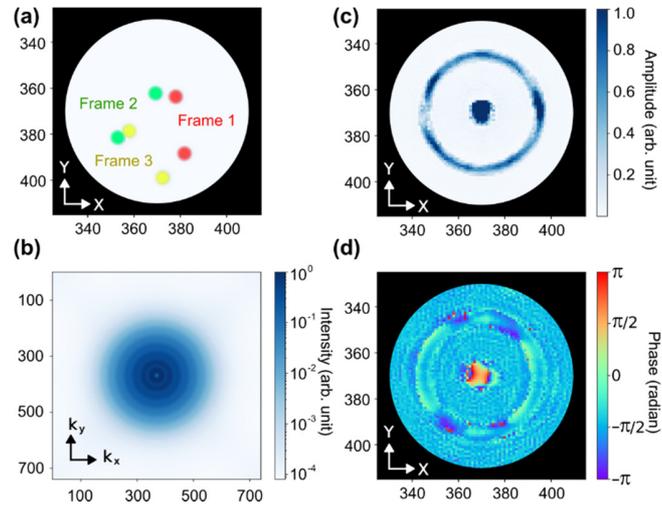

**Figure 4: CIDI of randomly oriented polaron pairs**. a) Image of a pair of polarons with fluctuating mean position and arbitrary orientation in two dimension is shown for three frames. The three different locations are represented by three different colours. b) Fluctuation spectrum of the pair of polarons with arbitrary orientation retrieved using CIDI. c) Amplitude and d) phase image recovered using CDI from the pattern in panel b. A central peak and outer ring are recovered; the width of the central peak and radial width of the outer ring match the size of the polarons, while the relative phase of π corresponds to the relative phase of the two polarons. The seperation of the polarons matches the radius of the ring.

**Extracting the properties of domain walls**

Finally we consider the case of fluctuating domain walls[33]. Here we simulate the case of metallic domain walls forming at random locations between otherwise insulating domains[34], but these results are indicative of other important cases, for instance ferromagnetic domain walls between antiferromagnetic domains[35]. In this case the fluctuation takes the form of a stripe with a strong aspect ratio corresponding to the (generally small) width of the domain wall and the (generally long) length of the insulating domain. In the case where the stripes are oriented, for instance along a particular crystal axis[16], it is possible to directly reconstruct the structure and recover the approximate size and shape of the domain walls. However, here we will consider the more general case of randomly oriented and positioned domain walls, as illustrated in Figure 5a. In this case CDI reconstruction fails to return any relevant information. Nevertheless, CIDI analysis alone is able to return two important parameters about the domain walls: their spectral dependence (and thus metallic character) and their approximate width.

We first show the retrieved fluctuation spectrum at one particular photon energy in which there exists contrast between the metallic domain wall and the insulating domains in Figure 5b. For these simulations we average over

10000 frames. The pattern is radially symmetric due to the random orientation of the domain walls. However, we can see the radial distribution is relatively complex due to the DC offset Airy disk effect discussed earlier. In principle if one could uniquely assign the DC changes to the fluctuations, for instance in a static measurement where we cooled to a point where domain walls just begin to form, this term could be fit and removed. In general however, this pattern overlaps strongly with the scattering from the long-axis of the domain walls (vertical in Fig. 5a) and prevents quantitative information being extracted. The high $k$ tails of the distribution, however, correspond to the thin width of the domain walls, and give a qualitative measure of this value. Such information is very difficult to extract from conventional measurements of fluctuations, which usually return the size of the normal, partially ordered background phase[18,20,36] and no information on the walls themselves.

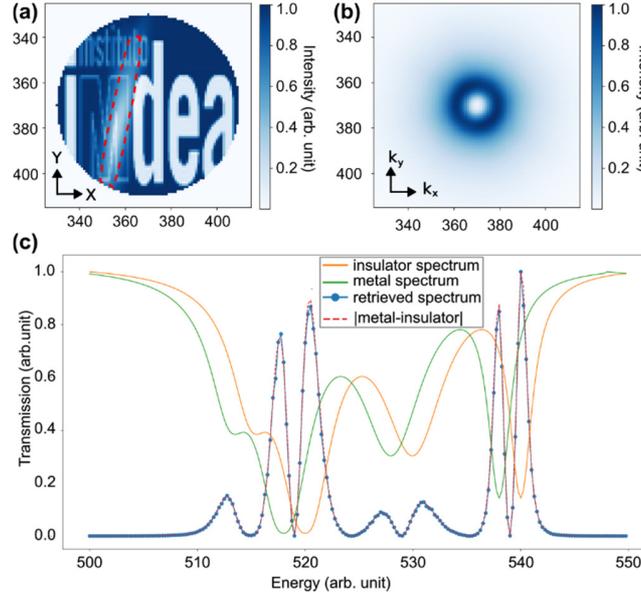

**Figure 5: CIDI spectroscopy of fluctuating domain walls**. a) Single snapshot of the sample shows presence of the metallic domain (highlighted by red dotted line) wall at arbitrary position and arbitrary orientation at a photon energy of 517.5 (arb. units). b) The CIDI isolated fluctuation spectrum showing DC suppression and circular symmetry. c) Absorption spectra of the insulating background, metallic domain walls, and CIDI retrieved absorption spectrum of the fluctuations. The CIDI absorption spectrum corresponds to the absolute difference of the metallic walls and insulating background.

We next turn to the spectral dependence of the fluctuations and the ability to extract the metallic nature of the domain walls[5]. We introduce an absorption spectrum for the insulating phase and for the metallic phase, here characterized by a simple shift to lower photon energies for the metallic phase. We then repeat the simulation varying the transmission of the two phases according to these two spectra, and recover the fluctuation spectrum using CIDI at each photon energy. We note a new dataset is generated for each photon energy, and so the spatial behaviour of the fluctuations is uncorrelated between each photon energy. We then sum the recovered fluctuation spectrum over all $k$ and plot the resulting transmission vs. photon energy. As can be seen, we recover that the spectrum of the fluctuations is the difference between the insulating background and metallic fluctuations. This is in sharp contrast to the average over the whole, pre-CIDI diffraction pattern, which instead returns a weighted average of the insulating and metallic spectra with a simple small shift to lower energies, obscuring the nature of the underlying metallic domain walls.

**Partial coherence and other challenges**

We now briefly discuss the effect of various imperfections in the experimental implementation on the CIDI analysis. Real experiments are characterized by partial coherence of the source and noise, which have been neglected in our simulations until this point.

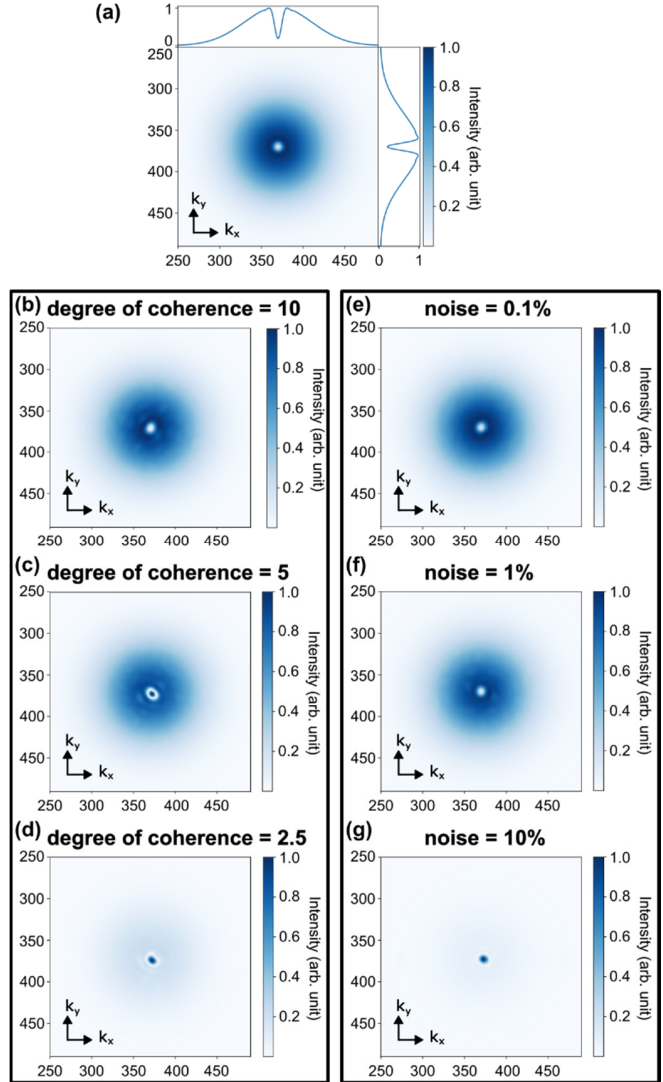

**Figure 6: Effect of noise and partial coherent source.** a) Fluctuation spectrum of a single vortex with fully coherent source and without noise is simulated. (b-d) Fluctuation spectrum in presence of partial coherence, with degree of coherence of b) 10, c) 5, and d) 2.5. (e-g) Fluctuation spectrum in presence of white noise level e) 0.1 %, f) 1 %, and g) 10 % of the maximum of the fluctuation spectrum.

The partial coherence of the source can be quantified using a Gaussian coherence function characterized by a coherence length $l_{coh}$[30,37,38]. By convoluting the Fourier transform of the coherence function with the otherwise fully coherent diffraction pattern $|FFT(E(r,t))|^2$, we obtain the partially coherent diffraction pattern of the object. To remain dimensionless, we consider the degree of coherence, defined by the ratio ($l_{coh}/L$) of the coherence length and the size of the object $L$, here corresponding to the maximum radial distance of the holographic reference apertures[37]. Fluctuation spectrum of a single vortex (figure 6a) is simulated for different degrees of coherence (Fig 6b-c). For high degrees of coherence the reconstruction remains of very high quality, but as the coherence drops a DC peak begins to appear. This is because the primary effect of limited coherence is to suppress the $R_i(k)R_j^*(k)$ terms, causing an inflation in the normalization of $|x(k)|^2$ in equation 6, and thus the DC term to equation 5 to cancel incorrectly. The exact impact of this will depend sensitively on the relative weights of the deterministic and stochastic contributions, but in general degrees of coherence much greater than 1 should be used in order to ensure good normalization. Alternatively scaling factors could be introduced into equation 5 to correct for this factor, ensuring that the DC component of the fluctuation spectrum is zero. As expected, CIDI thus relies heavily on the high spatial coherence of the source, a requirement which can be satisfied at 4th-generation synchrotron[39] or X-ray free electron laser sources[40].

We further explore CIDI in the presence of noise. In figures 6e-g we add white noise to the total diffraction pattern $|FFT(E(r,t))|^2$ and see the effect on the recovered fluctuation spectrum. We consider different amplitudes of noise, defined with respect to the maximum of the fluctuation spectrum without noise (figure 6a). At low noise levels (figure 6e, f) the recovered pattern is relatively unperturbed, but at higher noise levels a DC peak appears instead. Both this and the coherence measurements show the importance to CIDI of retrieving accurately the $R_i(k)R_j^*(k)$ normalization terms, which in turn requires high coherence and good signal to noise.

## Conclusions

We have demonstrated a new methodology, CIDI, capable of separating the deterministic and stochastic contributions to a coherent scattering pattern. We show that, from averaged diffraction pattern of multiple snapshots, it is possible to isolate the stochastic part through a FTH-like analysis. Because the stochastic contribution is necessarily confined to a restricted aperture it becomes possible to apply a CDI analysis[29] and return a mixture of qualitative and quantitative information regarding the real space behaviour of these fluctuations. Furthermore CIDI can be combined with X-ray spectroscopy[5] or other multidimensional studies to isolate the properties of the fluctuations from the deterministic variation of the sample in question. Although holographic apertures are used to provide phase stable references, the spatial resolution that can be obtained is set by the maximum-q captured, and the size of the holographic apertures only weakly affects the returned pattern via a discrete numbers of divergences in the recovered spatial pattern set by the zero of the corresponding Airy disk[26]. We note that while here we have focused exclusively on the stochastic portion, the same analysis could be used to remove the stochastic contribution and improve CDI reconstruction of the deterministic part of the signal. CIDI also acts as a *diagnostic* for the presence of stochastic contributions, and could be used to validate the interpretation of previous experiments[10,11].

We have demonstrated CIDI in three representative test cases – uncorrelated point-like defects (vortices), polaron-like pairs, and metallic domain walls in an insulating matrix – but there are many more examples of fluctuations at the nanoscale available where CIDI could be applied. For time resolved studies CIDI could shed light if signals observed in recent imaging experiments[10,11] are truly due to homogeneous melting of domains or due to stochastic domain changes. However, the use of CIDI to studying fast fluctuations does not actual require the use of femtosecond X-ray pulses; the limitation will be given by the coherence time of the light[41], which determines over what time window scattering contributions can add coherently at the detector. This means it may be possible to image femtosecond fluctuations using broadband continuous wave radiation, for instance the pink-beam of a synchrotron. In the future we foresee CIDI applied to a wide range of challenges where existing methods struggle.

## Author Contributions

ASJ conceived the study and performed the initial analytic calculations. AS performed the simulations and analysis of the results. Both authors contributed to writing of the manuscript.

## Conflicts of interest

There are no conflicts to declare.

## Acknowledgements

This work was funded by the Spanish AIE (projects PID2022-137817NA-I00 and EUR2022-134052). ASJ acknowledges the support of the Ramón y Cajal Program (Grant RYC2021-032392-I). IMDEA Nanociencia acknowledges support from the "Severo Ochoa" Programme for Centers of Excellence in R&D (MICIN, CEX2020-001039-S).